\documentclass[showpacs,prd,nofootinbib]{revtex4}
\usepackage{amssymb}
\usepackage{dcolumn}

\usepackage{graphicx}
\usepackage{dcolumn}
\usepackage{bm}

\begin{document}


\title{On the Constraint Equations in Einstein-aether Theories and the Weak Gravitational Field Limit }

\author{David Garfinkle}
\email{garfinkl@oakland.edu}
\affiliation{Department of Physics, Oakland University, Rochester, MI 48309, and \\
Michigan Center for Theoretical Physics, Randall Laboratory of Physics, University of Michigan, Ann Arbor, MI 48109}

\author{James Isenberg}
\email{isenberg@uoregon.edu}
\affiliation{Department of Mathematics and Institute for Theoretical Science, University of Oregon, Eugene, OR 
97403}

\author{Jose M. Martin-Garcia}
\email{jose@xact.es} 
\affiliation{Laboratoire Univers et Th\'eories, Observatoire de Paris, CNRS, Univ. Paris Diderot, 92190 Meudon, France, and\\
Institut d'Astrophysique de Paris, Univ. Pierre et Marie Curie, CNRS, 75014 Paris, France}

\begin{abstract}
We discuss the set of constraints for Einstein-aether theories, comparing the flat background case with what is expected when the gravitational fields are dynamic. We note potential pathologies occurring in the weak gravitational field limit for some of the Einstein-aether theories.

\end{abstract}
\pacs{04.20.Fy, 04.50.Kd}
\maketitle

\section{Introduction}

There has been significant interest in recent years in gravitational field theories that involve a fundamental vector field as well as the usual metric field. These include the paramerized family of Einstein-aether 
theories \cite{ted} and the TeVeS theory \cite{bekenstein}.  In addition, the Horava theory \cite{horava} can be viewed as a particular case of Einstein-aether theory, so it too is effectively  a gravitational field  theory with a fundamental vector field coupled to a metric field. A number of years ago, Isenberg and Nester \cite{jimandnester} noted a possible difficulty for certain field theories involving a vector field coupled to a spacetime metric field.\footnote{The same problem can arise for field theories involving tensor fields coupled to a metric field.} This difficulty involves the number of constraint equations the theory imposes on the choice of initial data for the theory, and how this number changes if one compares (i) the field theory with both the vector field and the metric field fully dynamic; and (ii) the field theory with the spacetime metric fixed and flat, and only the vector field fully dynamic.

To understand the problem, it is useful to first recall the situation for the Einstein-Maxwell field theory. On a fixed flat spacetime background, with no dynamical gravitational fields, the Maxwell theory imposes the two constraints $D_c E^c=0$ and $D_c B^c=0$ on the choice of the Maxwell initial data $\{E^a ,B^a\}$. For the fully dynamic Einstein-Maxwell theory, there are $2+4$ constraints---including the four Einstein constraints $R - K^{cd}K_{cd} + K^2 = \eta [ E^c E_c +B^c B_c ]$ and $D_c K^c_a - D_a K = \eta [E \times B]_a$  as well as the two Maxwell divergence constraints listed above --- restricting the choice of the Einstein-Maxwell data $\{ h_{ab}, K_{ab}, E^a, B^a\}$.\footnote{Here $E^a$ and $B^a$ are the electric and magnetic spatial vector fields,  $h_{ab}$ is the spatial metric field, $K_{ab}$ is the extrinsic curvature, $K$ is the trace 
of $K_{ab}$ (with respect to $h_{ab}$), $D_a$ is the covariant derivative compatible with $h_{ab}$, $R$ is the spatial scalar curvature, $\eta$ is a coupling constant, and lower case Latin indices range over the three spatial directions.} This (Einstein-Maxwell theory) is the normal situation for a well-behaved theory: there are $l=2$ constraints for the fixed background (Maxwell) theory, there are $l+4$ constraints in the dynamical gravitational field (Einstein-Maxwell) theory, and the constraints behave well in the weak gravitational field limit. 

If alternatively one considers the Einstein-vector field theory with the spacetime action principle\footnote{Here $U^{\alpha}$ is a spacetime vector field, $g_{\alpha \beta}$ is the spacetime metric field with the compatible spacetime covariant derivative $\nabla _\alpha$ and spacetime scalar curvature $\mathcal R$, $\rho$  is a coupling constant, $m$ is a (``mass") constant, and lower case Greek indices range over four spacetime directions.}
\begin{equation}
\label{U-action}
S[U^\alpha, g_{\alpha \beta}] = \int d^4x \sqrt{-g} \Big\{\mathcal{R} - \rho [ -\frac{1}{2}\nabla_{\alpha} U_{\beta} \nabla^{\beta} U^{\alpha} - \frac{1}{2} m^2 U_{\alpha} U^{\alpha}] \Big\}, 
\end{equation}
one finds the following alternative situation (See \cite{jimandnester} for details):  For the flat background theory, if one does a space + time decomposition $U^{\alpha} \to \{U^{\perp}, U^a\}$ of the fields, and if one defines the conjugate momenta $\{ \Pi_{\perp}, \Pi_a\}$, then there are six constraint equations on the initial data, of the following form
\begin{equation}
\Pi^a = D^a U^{\perp}
\end{equation}
\begin{equation}
D^b \Pi_{\perp} = D^b D_c U^c + m^2 U^b. 
\end{equation}
These constraints effectively reduce the theory to one (vector field) degree of freedom. However, for the fully dynamic Einstein-vector version of this theory, instead of having $6+4$ constraints on the initial data $\{ h_{ab}, K_{ab}, U^{\perp}, \Pi_{\perp}, U^a, \Pi_a \}$, one finds that there are only four constraints (analogous to, but much messier algebraically than the usual four Einstein constraints). There are then effectively four vector field degrees of freedom, plus two gravitational degrees of freedom. However, if the Einstein-vector theory is well behaved then its weak field limit must reduce to vacuum linearized gravity for the metric and the flat background theory for the vector field. Consequently, the mismatch between the 
number of constraints shows that the Einstein-vector theory behaves poorly as the gravitational field approaches flatness. 

There are two features of the Einstein-vector theory with action $S[U^\alpha, g_{\alpha \beta}]$ (as in (\ref{U-action}) above) which lead to this problem with constraints (and correspondingly with degrees of freedom) in the weak field limit: (i) the flat background version of the theory has constraints; and (ii) the theory is ``derivative-coupled" in the sense that the non-gravitational part of the action involves covariant derivatives. This first feature is crucial, since the difficulty of interest involves the ``loss" of constraints that occurs as the dynamical gravitational field is turned on. The second feature is also crucial, since it is the presence of terms involving $K_{ab}$ in the non-gravitational part of the action principle (following its $3+1$ decomposition) that leads to the loss of constraints. We note that the Einstein-Maxwell theory has constraints in its flat background version, but is not derivative-coupled; consequently it avoids this problem. The Einstein-massive Klein-Gordon theory, whose action is the same as $S[U^\alpha, g_{\alpha \beta}]$ above  \emph{except} that the term $-\frac{1}{2}\nabla_{\beta} U_{\alpha} \nabla^{\beta} U^{\alpha}$ replaces the term $-\frac{1}{2}\nabla_{\alpha} U_{\beta} \nabla^{\beta} U^{\alpha}$ in (\ref{U-action}), is derivative-coupled, but has no constraints in its flat background version; consequently it too avoids the problem.

Where do these constraints in the flat space theory come from?  The easiest way to see this is to consider the process of finding a Hamiltonian formulation of the theory.  Recall that to find a Hamiltonian formulation of a theory, one begins with
a Lagrangian and varies the Lagrangian with respect to the velocities to obtain the momenta.  The next step is to invert the
velocity-momentum relation to obtain the velocity in terms of the momentum.  However, that step can fail if the velocity-momentum relation is not invertible.  Such a lack of invertibility gives rise to constraints, and it is precisely this situation that gives rise to the constraints found in the flat space theories studied in \cite{jimandnester}.

What about the Einstein-aether theories? As we discuss below in Section \ref{Eatheories}, these theories form a four-parameter set, with (essentially) every theory in the set involving derivative-coupling. Hence, the first step in applying the methods of \cite{jimandnester} to the Einstein-aether theories is to determine which of the theories, in their flat background versions, involve constraint equations. We do this in Section \ref{HamDir}, working with the theories in Hamiltonian form.  We find that some of the flat background theories involve no constraints (``safe" theories), some of them involve constraints regardless of the initial data (``endangered" theories) and the rest involve constraints for some ranges of data, but not for others (``conditionally endangered" theories). In Section \ref{PrefTh} we consider the implications of the results of 
Section \ref{HamDir} for Einstein-aether theories.  This is not as straightforward as for the vector theories of 
\cite{jimandnester} because the weak field limit of Einstein-aether theories is \emph{not} the flat background theory, but 
rather a theory that still couples the vector field to the metric perturbation.  Nonetheless we find that the theories that we call ``endangered'' really are pathological in the weak field limit.  We also consider the possible implications of the 
``conditionally endangered'' theories.

\section{Einstein-aether Theories}
\label{Eatheories}

The Einstein-aether theories make up a four-parameter family of classical metric-vector theories specified by the spacetime action principles 
\begin{equation}
\label{Ea-action}
S_{Ea}[u^\alpha, g_{\alpha \beta}] = \int d^4x \sqrt{-g} \Big\{\mathcal{R} - K^{\alpha \beta}_{\mu \nu} \nabla_{\alpha} u^{\mu} \nabla_{\beta} u^ {\nu} + \lambda (1 + g_{\alpha \beta} u^{\alpha} u^{\beta}) \Big\},
\end{equation}
where our conventions on the spacetime metric and  indices are as above, where $u^{\alpha}$ denotes a spacetime vector field, where $\lambda$ is a Lagrange multiplier, and where the aether-parameter matrix $K^{\alpha \beta}_{\mu \nu}$ takes the form
\begin{equation}
K^{\alpha \beta}_{\mu \nu} = c_1 g^{\alpha \beta}g_{\mu \nu} + c_2 \delta ^{\alpha}_{\mu} \delta ^{\beta}_{\nu} + c_3 \delta ^{\alpha}_{\nu} \delta ^{\beta}_{\mu} - c_4 u^{\alpha} u^{\beta} g _{\mu \nu}.
\label{aetherL}
\end{equation}
with (constant) parameters $\{ c_1, c_2, c_3, c_4\}$. We note that if $c_2=c_3=c_4=0$ and $\lambda = 0$, then this is the Einstein-Klein-Gordon (massless) theory; if $c_1=c_2=c_4=0$ and $\lambda = 0$, then we have essentially the Einstein-vector theory (\ref{U-action}) discussed above. If $c_1+c_3=c_2=c_4=0$ and $\lambda$ does not vanish, then this is the particular version of the Einstein-aether theory discussed in \cite{clayton}.

The presence of the Lagrange multiplier term serves to enforce the \emph {a priori} restriction 
\begin{equation}
\label{restriction}
g_{\alpha \beta} u^{\alpha} u^{\beta}=-1, 
\end{equation} 
which says that the vector field must always be a unit-length timelike vector field. While this restriction is crucial to the physical application of the Einstein-aether theories, it does not appear to have a major qualitative effect on the issue under discussion here. 

As noted above, our focus here is on the Einstein-aether theories with a fixed flat background. In that case, with $\eta_{\alpha \beta}$ representing the Minkowski metric, the action reduces to 
\begin{equation}
\label{Eaflat-action}
S_{Ea(flat)}[u^\alpha] = \int d^4x \Big\{- K^{\alpha \beta}_{\mu \nu} \nabla_{\alpha} u^{\mu} \nabla_{\beta} u^ {\nu} + \lambda (1 + \eta_{\alpha \beta} u^{\alpha} u^{\beta}) \Big\},
\end{equation}
the aether-parameter matrix takes the form 
\begin{equation}
K^{\alpha \beta}_{(flat) \mu \nu} = c_1 \eta^{\alpha \beta}\eta_{\mu \nu} + c_2 \delta ^{\alpha}_{\mu} \delta ^{\beta}_{\nu} + c_3 \delta ^{\alpha}_{\nu} \delta ^{\beta}_{\mu} - c_4 u^{\alpha} u^{\beta} \eta _{\mu \nu}, 
\label{aetherLflat}
\end{equation}
and the unit length condition (\ref{restriction}) can be written as 
\begin{equation}
\label{flatrestriction}
\eta_{\alpha \beta} u^{\alpha} u^{\beta}=-1. 
\end{equation} 

One way to determine the presence of constraint equations in the initial value formulation of the (flat background) aether theories is to first obtain the spacetime covariant field equations for $u^{\alpha}$ (by varying the action $S_{Ea(flat)}[u^\alpha]$), then rewrite these equations in $3+1$ form, and finally search among the $3+1$ equations for explicit constraints. The Hamilton-Dirac approach provides a more systematic way to find the constraints. Since we are primarily concerned here with determining for which of the four-parameter family of aether theories there are constraints (rather than determining the exact nature of these constraints),  in fact we need only carry out the first part of the Hamilton-Dirac analysis, as we discuss in the next section. For completeness, we summarize the covariant approach in the Appendix.

\section{Hamilton-Dirac analysis}
\label{HamDir}

The first step in the Hamilton-Dirac analysis of a given field theory is to carry out a $3+1$ (space + time) decomposition of the fields and their derivatives, and to substitute the resulting expressions into the Lagrangian for the given theory. For the present case, 
 we choose a standard slicing for the flat background spacetime with $n^{\alpha}$ as the unit time-like normals, and introduce the tensor 
\begin{equation}
{h_{\alpha \beta}} := {g_{\alpha \beta}} + {n_\alpha}{n_\beta} \;  .
\end{equation}
We note that spacetime tensors which are orthogonal to $n^\alpha$ can also be considered as spatial tensors.  Viewed in this
way, $h_{\alpha \beta}$ (which we can also write as $h_{ab}$) is both the spatial metric and a projection operator that takes spacetime tensors to their 
spatial part.  Using projection with respect to $h_{\alpha \beta}$ and contraction with $n^\alpha$ we can decompose
any spacetime tensor into spatial tensors.  In particular, the four vector  $u^\alpha$ can be  decomposed as 
\begin{equation}
{u^\alpha} = V {n^\alpha} + {w^\alpha} ,
\label{udecomp}
\end{equation}
where $V$ is a scalar and $w^\alpha$ (which we can also write as $w^a$) is a spatial vector.  
The unit-length condition (\ref{flatrestriction}) then implies that 
$V$ is not an independent field but is
instead determined by $w^a$ through the relation
\begin{equation}
{V^2} = 1 + {w^a}{w_a} .
\label{Vsq}
\end{equation}
Similarly, we can decompose the spacetime derivative of $u^{\alpha}$ as 
\begin{equation}
{\nabla _\alpha} {u_\beta} = - {n_\alpha}{n_\beta}{\dot V} + {n_\beta}{D_\alpha}V - {n_\alpha}{{\dot w}_\beta}  
+{D_\alpha}{w_\beta} \:,
\label{gradu}
\end{equation}
where an overdot denotes the derivative with respect to time, and $D_\alpha$ (which we can also write as $D_a$) denotes the 
spatial covariant derivative operator.  

Using eqn.\ (\ref{gradu}) in eqn.\ (\ref{Eaflat-action}) we find that the Lagrangian corresponding to the action $S_{Ea(flat)}[u^\alpha] $    takes the form
\begin{eqnarray}
L = M {{\dot w}^a}{{\dot w}_a} - (M + {c_{23}}){{\dot V}^2}
+ 2 {{\dot w}^a} ( - {c_3}{D_a}V 
+ {c_4}V {w^b}{D_b}{w_a})
\nonumber
\\
- 2 {\dot V} ({c_2} {D_a}{w^a} + {c_4}V {w^a}{D_a} V) + Z \:.
\label{lagrangian2}
\end{eqnarray}
Here the quantities $M$ and $Z$ are given by
\begin{eqnarray}
M := {c_1} + {c_4}{V^2},
\label{Mdef}
\\
Z := {c_1}({D_a}V{D^a}V - {D_a}{w_b}{D^a}{w^b}) - {c_2} {{({D_a}{w^a})}^2}
- {c_3} {D_a}{w_b}{D^b}{w^a} 
\nonumber
\\
+ {c_4} (({w^a}{D_a}{w_b})({w^c}{D_c}{w^b}) -
{{({w^a}{D_a}V)}^2}),
\end{eqnarray}  
and we are using the notation of \cite{ted} in which $c_{23}$ is an abbreviation for ${c_2}+{c_3}$ (with corresponding abbreviations 
for any other sums of the $c_i$).

While the condition (\ref{Vsq}) is a constraint, it is one which (if we restrict attention to non-negative $V$) can be eliminated algebraically. More specifically, we adopt the point of view that (flat background) aether theory is a theory of the spatial
vector field $w^a$,  and that any terms in the Lagrangian that depend on $V$ are
to be viewed as simply more complicated functions of $w^a$ given by substituting
$\sqrt {1 + {w^a}{w_a}}$ for each occurrence of $V$.  In this way, 
the aether theory Lagrangian has no \emph{a priori} constraints and correspondingly no need for the  Lagrange multiplier  term.  In particular we may substitute 
\begin{equation}
{\dot V} = {V^{-1}}{w_a}{{\dot w}^a}
\label{Vdot}
\end{equation}
along with $V= \sqrt {1 + {w^a}{w_a}}$ into the Lagrangian expression (\ref{lagrangian2}),
thereby obtaining
\begin{equation}
L = {q_{ab}} {{\dot w}^a}{{\dot w}^b} + 2 {{\dot w}^a}{B_a} + Z, 
\label{lagrangian3}
\end{equation}
where the tensor $q_{ab}$ and vector $B_a$ are given by
\begin{eqnarray}
{q_{ab}} := M {h_{ab}} - (M + {c_{23}}){V^{-2}}{w_a}{w_b},
\label{qdef}
\\
{B_a} := {c_4}{V^2} {w^b}{D_b}({V^{-1}}{w_a}) - {c_3} {D_a}V - {c_2} {V^{-1}}{w_a}{D_b}{w^b}.
\end{eqnarray}
This is the $3+1$ form of the Lagrangian expression for the aether theories which we work with here. 

The next step of the Hamilton-Dirac analysis is to calculate the expressions for the momenta $p_a$ conjugate to the fields $w^a$ as functions of $w^a$ and $\dot w^a$, and then attempt to invert these expressions to obtain new expressions for $\dot w^a$ as functions of $w^a$ and $p_a.$ It is at this step---a key step in the specification of the Legendre transform which maps from a Lagrangian formulation to a Hamiltonian formulation of a given theory---that one finds out if the theory has any constraints on the choice of initial data sets. Such constraints exist if and only if one \emph{cannot} invert the map from $p_a(w, \dot w)$ to $\dot w^a(w,p)$. 

To determine the full explicit set of constraints for a given theory via the Hamilton-Dirac analysis, one proceeds from this step to construct the Hamiltonian (or set of Hamiltonians) for the theory, incorporating the constraints obtained from the non-invertibility just discussed, and one calculates the time derivatives of the constraints using this Hamiltonian. We leave the remaining details to other references (see, e.g., \cite{Dirac}). Here, our only real concern is to determine the sets of choice of the Einstein-aether parameters $\{c_1,c_2, c_3, c_4\}$ for which there are constraints and those choices for which there are not.

Calculating the conjugate momentum $p_a$ by varying the Lagrangian (\ref{lagrangian3}) with respect to ${\dot w}^a$,  
\begin{equation}
{p_a}\delta {{\dot w}^a} = \delta L = \delta {{\dot w}^a} 2 ({q_{ab}} {{\dot w}^b} + {B_a}), 
\end{equation}
we obtain 
\begin{equation}
{p_a} =  2 ({q_{ab}} {{\dot w}^b} + {B_a}).
\label{momentum}
\end{equation}
From eqn.\ (\ref{momentum}), one immediately sees that the momentum-velocity relation can be inverted 
if and only if the matrix $q_{ab}$ is invertible.  It is easy to see that if $q_{ab}$ is not invertible then there is a constraint,
since for any vector $s^a$ for which ${q_{ab}}{s^b}=0$ it follows from eqn.\ (\ref{momentum}) that
\begin{equation}
{s^a}({p_a}-2{B_a})=0  .
\end{equation}

To find the conditions under which $q_{ab}$ is invertible, we note (from eqn.\ (\ref{qdef})), 
that any vector orthogonal to $w^a$ is necessarily an  eigenvector
of $q_{ab}$,  with eigenvalue $M$.  Furthermore, it also follows from eqn.\ (\ref{qdef}) that  $w^a$ is generally an eigenvector of $q_{ab}$ with eigenvalue ${V^{-2}}N$, where
\begin{equation}
N = M - {c_{23}}{w_a}{w^a}.
\label{Ndef}
\end{equation}
Therefore $q_{ab}$ is invertible if and only if $MN \ne 0$.  This result immediately gives rise to a classification of (flat background) aether
theories,  depending on whether  $MN$ vanishes at all points of configuration space (``endangered" theories), at no points
of configuration space (``safe" theories), or at only some points of configuration space (``conditionally endangered" theories).
We would like to express this classification directly in terms of the parameters of the theory.  From eqns.\ (\ref{Mdef})
and (\ref{Ndef}) we determine  that 
\begin{equation}
MN = ({c_{14}} + {c_4}{w_a}{w^a})({c_{14}}+({c_4}-{c_{23}}){w_a}{w^a}).
\end{equation}
Consequently, we find that  the \emph{endangered theories} satisfy either the condition 
\begin{equation}
{c_1} = {c_4} = 0, 
\label{unsafe1}
\end{equation}
or the condition
\begin{equation}
{c_1} \ne 0 , \; \; {c_4} = {c_{23}} = - {c_1}.
\label{unsafe2}
\end{equation}
In particular, the theories studied in 
\cite{jimandnester} are analogs for non-unit massive fields of endangered aether theories.  
The \emph{safe theories} satisfy the conditions
\begin{equation}
{c_{14}} \ne 0 , \; \; \; {\frac {c_4} {c_{14}}} \ge 0 , \; \; \; {\frac {{c_4}-{c_{23}}} {c_{14}}} \ge 0 .
\label{safe}
\end{equation}
Any theory that is neither endangered nor safe is conditionally endangered.  This completes our classification.

For the safe theories, it is straightforward to proceed to construct a Hamiltonian. In particular, so long as 
$MN \ne 0$,  we calculate  from eqn.\ (\ref{qdef}) that
\begin{equation}
{{({q^{-1}})}^{ab}} = {\frac 1 {MN}} (N{h^{ab}} + (M+{c_{23}}){w^a}{w^b}).
\label{qinv}
\end{equation}
It then  follows that the velocity-momentum relation can be inverted to yield
\begin{equation}
{{\dot w}^a} = {\textstyle {\frac 1 2}}{{({q^{-1}})}^{ab}} ({p_b} - 2 {B_b}).
\label{velocity}
\end{equation}
Using the non-constrained definition of the Hamiltonian, we have 
\begin{equation}
H = {p_a}{{\dot w}^a} - L = {\textstyle {\frac 1 4}}{{({q^{-1}})}^{ab}} ({p_a} - 2 {B_a}) ({p_b} - 2 {B_b}) - Z.
\end{equation}
Then using eqn.\ (\ref{qinv}), we obtain
\begin{equation}
H = {\frac 1 {4MN}} (N{h^{ab}} + (M+{c_{23}}){w^a}{w^b})({p_a} - 2 {B_a}) ({p_b} - 2 {B_b}) - Z.
\end{equation}

A Hamiltonian which incorporates the collateral constraints can also be constructed for the endangered and conditionally endangered theories; the details of this construction (peripheral to our concerns  here) are discussed in \cite{Dirac}.

\section{Conclusions}
\label{PrefTh}

What are the possible physical implications of our classification of (flat background) aether theories?  In \cite{jimandnester} it is shown that
the extra constraints in the flat space vector theories studied in that paper are not present in the gravitating 
versions of those theories, as a consequence of the presence of many terms containing $\dot h_{ab}$ appearing throughout the $3+1$ expression for the Lagrangian of the coupled theories. 
We have not performed the corresponding analysis for the Einstein-aether theories, endangered or otherwise. Their Lagrangians do, however, contain a similar number of $\dot h_{ab}$ terms.  We note in addition that it has been shown that the diffeomorphism invariance of Einstein-aether theories gives rise in each of them to at least four constraints, just as in general 
relativity.\cite{tedconstraint,seifert}. 
We are consequently led 
to believe that in general the Einstein-aether theories have four (and no more) constraints.  However, unlike the case of 
\cite{jimandnester} even if we were to show that the Einstein-aether theories have exactly four constraints, that would
not allow us to conclude that the weak field limit of the endangered theories is pathological, because the weak field limit of 
Einstein-aether theories is not the non-gravitating aether theory, but still contains a residual coupling between the 
vector field and the metric perturbation.  Nonetheless, the presence of constraints in the flat space endangered theories at least leads to the suspicion that there is something wrong with the corresponding Einstein-aether theories.  And that suspicion turns out to be justified.  The endangered theories all have ${c_{14}}=0$.  (see eqns.\ (\ref{unsafe1})-(\ref{unsafe2})).  
However, as noted in \cite{tedreview} Einstein-aether theories with ${c_{14}}=0$ have infinite speed for the spin 0 and spin 1
modes of the theory.  Thus the endangered flat space theories serve as a diagnostic of a pathology in the corresponding gravitating theory.  

What about the conditionally endangered theories?  Here the number of constraints changes at particular ``bad'' points of 
configuration space.  Certainly, this should lead one to worry about the physical viability of the flat space theory itself,
since even if one were to restrict initial data to a ``good'' region of configuration space (where no constraints are present)
there remains the possibility of the dynamics causing the
system to evolve to a ``bad'' part of configuration space, thus resulting in an ill-defined or singular evolution.  However,
it is not clear that these properties of the flat space theory give any cause to worry about the gravitating theory: note that 
the bad points of configuration space typically have ${w^a} \sim 1$ and are thus far from the weakly gravitating case.  Thus
the pathologies of the conditionally endangered theories cannot be considered as a reliable guide to the behavior of the 
Einstein-aether theory with the same $c_i$.  Nonetheless, a non-gravitating theory with singular evolution may give rise to naked singularities when coupled to gravity.  This raises the question of whether gravitational collapse gives rise to naked
singularities in some of the Einstein-aether theories.  This question could be addressed numerically using e.g. the methods of
\cite{meandted}

\begin{acknowledgments}
We would like to thank Richard Woodard, John Moffat, and especially Ted Jacobson for helpful discussions.
This work was supported in part by the National Science Foundation
under grants PHY-0855532 and PHY-1205202 through Oakland University and under grant PHY-0968612
through the University of Oregon, and by the French ANR Grant BLAN07-1\_201699.
\end{acknowledgments}

\appendix
\section{Covariant field equations}

Varying the Lagrangian of eqn. (\ref{aetherL}) with respect to the aether field $u^\alpha$ yields the field equation
\begin{equation}
{\nabla _\alpha}{{J^\alpha}_\beta} + \lambda {u_\beta} + {c_4} {a_\alpha}{\nabla _\beta}{u^\alpha} = 0,
\label{aethermotion}
\end{equation}
where ${{J^\alpha}_\beta} = {{K^{\alpha \mu}}_{\beta \nu}}{\nabla _\mu}{u^\nu}$ and ${a_\beta}
={u^\alpha}{\nabla _\alpha}{u_\beta}$.  
Contracting eqn. (\ref{aethermotion}) with $u^\beta$ and solving for $\lambda$ yields
\begin{equation}
\lambda = {u^\beta}{\nabla _\alpha}{{J^\alpha}_\beta} + {c_4}{a^\alpha}{a_\alpha}
\end{equation}
and substituting this into eqn (\ref{aethermotion}) yields
\begin{equation}
0= {\nabla _\alpha}{{J^\alpha}_\beta} + {c_4} {a_\alpha}{\nabla _\beta}{u^\alpha} + 
{u_\beta} ( {u^\gamma}{\nabla _\alpha}{{J^\alpha}_\gamma} + {c_4}{a^\alpha}{a_\alpha} ) .
\end{equation}
However, we have 
\begin{equation}
{{J^\alpha}_\beta} = {c_1} {\nabla ^\alpha}{u_\beta} + {c_2} {{\delta ^\alpha}_\beta} {\nabla _\gamma}{u^\gamma}
+ {c_3} {\nabla _\beta} {u^\alpha} - {c_4} {u^\alpha}{a_\beta}.
\end{equation}
So after some straightforward but tedious algebra the field equation becomes
\begin{eqnarray}
0 = {c_1} [ {\nabla ^\alpha} {\nabla _\alpha} {u_\beta} - {u_\beta} {\nabla ^\alpha}{u^\gamma}
{\nabla _\alpha}{u_\gamma} ] + {c_{23}} [ {\nabla _\beta}{\nabla _\gamma}{u^\gamma} + {u_\beta}
{u^\gamma}{\nabla _\gamma}{\nabla _\alpha}{u^\alpha} ] 
\nonumber
\\
+ {c_4} [ {a_\alpha}{\nabla _\beta}{u^\alpha} + 2 {u_\beta} {a^\alpha}{a_\alpha} - {u^\alpha}{\nabla _\alpha}{a_\beta}
-{a_\beta}{\nabla _\alpha}{u^\alpha} ].
\label{fieldeqn}
\end{eqnarray}

We now want to produce an evolution equation for $w^a$ by substituting the decomposition of $u^\alpha$ of
eqn. (\ref{udecomp}) into eqn. (\ref{fieldeqn}).
At first it might then seem that the field equations might be an overdetermined
system, since they seem to provide four equations for the three components of $w^a$.
However, the field equation vector is automatically orthogonal to $u^\alpha$ so
it  suffices to impose the spatial projection of the field equations since
spatial part of the field equations along with orthogonality to $u^\alpha$ 
implies time part of the field equations.  Some straightforward but tedious calculation then shows that the spatial
projection of eqn (\ref{fieldeqn}) becomes
\begin{equation}
M {{\ddot w}_b} - {c_{23}} V {\ddot V} {w_b} = {R_b} ,
\label{motion1}
\end{equation}
where $M$ is given by eqn. (\ref{Mdef}) and $R_b$ is given by
\begin{eqnarray}
{R_b} = {c_1} [ {D^a}{D_a}{w_b} - {w_b} ( {{\dot V}^2} - {D^a}V{D_a}V 
-{{\dot w}^a}{{\dot w}_a} + {D_a}{w_c}{D^a}{w^c}) ]
\nonumber
\\
+ {c_{23}} [ {D_b} {\dot V} + {D_b}{D_a}{w^a} + {w_b} ( V {D_a}{{\dot w}^a} +
{w^a}{D_a} ({\dot V} + {D_c}{w^c}))]
\nonumber
\\
+ {c_4} [ - ({\dot V} + {D_a} {w^a} ) {L_b} - V {\dot V} {{\dot w}_b} 
- V {{\dot w}^a}{D_a}{w_b} - V {w^a}{D_a} {{\dot w}_b} 
\nonumber
\\
- {w^a}{D_a}{L_b} - (V{\dot V} + {w^a}{D_a}V){D_b} V +
{L^a}{D_b}{w_a} + 2 {w_b} ({L^a}{L_a} - 
{{(V{\dot V} + {w^a}{D_a}V)}^2} ) ].
\end{eqnarray}   
Here the vector $L_a$ is the spatial projection of the acceleration vector $a_\alpha$.
A straightforward computation yields
\begin{equation}
{L_a} = V {{\dot w}_a} + {w^b}{D_b} {w_a}.
\end{equation}

The field equation is not quite in the form that we would like, namely
with ${\ddot w}_a$ alone on one side of the equation and no quantities with second time
derivatives on the other side of the equation.  However, we can easily
put it in that form by using eqn. (\ref{Vsq}) and its time derivatives.  In particular, the first time derivative
of eqn. (\ref{Vsq}) yields eqn. (\ref{Vdot}) while a second time derivative yields
\begin{equation}
{\ddot V} = {V^{-1}} [ {w^a}{{\ddot w}_a} + {{\dot w}^a}{{\dot w}_a} - {{\dot V}^2} ].
\label{ddotV}
\end{equation}
Then contracting eqn (\ref{motion1}) with $w^a$ and using eqn. (\ref{ddotV}) we find
\begin{equation}
N V {\ddot V} = {R_a}{w^a} + M ({{\dot w}_a}{{\dot w}^a} - {{\dot V}^2}),
\end{equation} 
where the quantity $N$ is given by eqn. (\ref{Ndef}).
Then on substituting this result into eqn (\ref{motion1}) we obtain
\begin{equation}
M N {{\ddot w}_a} =  N {R_a}  + 
{c_{23}} [ {R_b}{w^b} + M({{\dot w}_b}{{\dot w}^b} - {{\dot V}^2}) ] {w_a}.
\label{motion2}
\end{equation}
We note that if $MN \ne 0$, eqn. (\ref{motion2}) is an equation of motion for $w^a$, while if
$MN=0$ it is a constraint equation.

\end{document}